\documentclass[twocolumn,showpacs,preprintnumbers,amsmath,amssymb,nofootinbib]{revtex4}

\usepackage{graphicx}
\usepackage{color}
\usepackage{dcolumn}
\usepackage{bm}

\newtheorem{proposition}{Proposition}

\newenvironment{result}{\ \\[0.5\baselineskip]\it}{\\[-0.5\baselineskip]}

\begin{document}

\preprint{APS/123-QED}

\title{Two-dimensional approach to relativistic positioning systems}

\author{Bartolom\'{e} Coll}
 \email{bartolome.coll@obspm.fr}
\affiliation{%
Syst\`emes de r\'ef\'erence relativistes, SYRTE-CNRS,\\
Observatoire de Paris, 75014 Paris, France.
}%

\author{Joan Josep Ferrando}
 \email{joan.ferrando@uv.es}
\author{Juan Antonio Morales}%
 \email{antonio.morales@uv.es}
\affiliation{%
Departament d'Astronomia i Astrof\'{\i}sica, \\Universitat de
Val\`encia, 46100 Burjassot, Val\`encia, Spain.
}%

\date{\today}

\begin{abstract}
A relativistic positioning system is a physical realization of a
coordinate system consisting in four clocks in arbitrary motion
broadcasting their proper times. The basic elements of the
relativistic positioning systems are presented in the
two-dimensional case. This simplified approach allows to explain and
to analyze the properties and interest of these new systems. The
positioning system defined by geodesic emitters in flat metric is
developed in detail. The information that the data generated by a
relativistic positioning system give on the space-time metric
interval is analyzed, and the interest of these results in
gravimetry is pointed out.

\end{abstract}


\maketitle

\section{Introduction}

In Astronomy, Space physics and Earth sciences, the increase in the
precision of space and time localization of events is associated
with the increase of a better knowledge of the physics involved. Up
to now all time scales and reference systems, although incorporating
so called `relativistic effects'\footnote{Today, this appellation
seems rather out of place. There are not `relativistic effects' in
relativity, as they are not `Newtonian effects' in Newtonian theory.
There are defects in the Ptolemaic theory of epicycles that could be
corrected by Newtonian theory, and there are defects in Newtonian
theory that may be corrected by the theory of relativity. But then
their correct appellation is not that of `relativistic effects' but
rather that of `Newtonian defects'.} in their development, start
from Newtonian conceptions.

In Relativity, the space-time is modeled  by a four-dimensional
manifold. In this manifold, most of the coordinates are usually
chosen in order to simplify mathematical operations, but some of
them, in fact a little set, admit more or less simple physical
interpretations. What this means is that some of the ingredients of
such coordinate systems (some of their coordinate lines, surfaces or
hypersurfaces) may be {\em imagined} as covered by some particles,
clocks, rods or radiations submitted to particular motions. But the
number of such {\em physically interpretable} coordinate systems
that can be {\em physically constructed} in practice is strongly
limited.

In fact, among the at present physically interpretable coordinate
systems, the only one that may be generically
constructed\footnote{Such a statement is not, of course, a {\em
theorem}, because involving real objects, but rather an {\em
epistemic} assertion that results from the analysis of methods,
techniques and real and practical possibilities of physical
construction of coordinate systems at the present time.} is the one
based in the Poincar\'e-Einstein protocol of
synchronization\footnote{The Poincar\'e-Einstein protocol of
synchronization is based in two-way light signals from the observer
to the events to be coordinated.}, also called {\em radar system}.
Unfortunately, this protocol suffers from the bad property of being
a retarded protocol (see below).

Consequently, in order to increase our knowledge of  the physics
involved in phenomena depending on the space-time localization of
their constituents\footnote{This includes, in particular, making
relativistic gravimetry.}, it is important to learn to construct
physically {\em good} coordinate systems of relativistic quality.

To clearly differentiate the coordinate system as a mathematical
object from its realization as a physical
object\footnote{Different physical protocols, involving different
physical fields or different methods to combine them, may be given
for a unique mathematical coordinate system.}, it is convenient to
characterize this physical object with a proper name. For this
reason, the physical object obtained by a peculiar materialization
of a coordinate system is called a {\em  location system}
\cite{coll-1, coll-tarantola, coll-3}. A location system is thus a
precise protocol  on  a particular set of physical fields allowing
to materialize a coordinate system.

A location system may have some specific properties \cite{coll-1,
coll-3}. Among them, the more important ones are those of being {\em
generic}, i.e. that can be constructed in any space-time of a given
class, (gravity-){\em free}, i.e. that the knowledge of the
gravitational field is not necessary to construct it\footnote{As a
physical object, a location system lives in the physical space-time.
In it, even if the metric is not known, such objects as point-like
test particles, light rays or signals follow specific paths which, a
priori, allow  constructing location systems.}, and {\em immediate},
i.e. that every event may know its coordinates without delay. Thus,
for example, location systems based in the Poincar\'e-Einstein
synchronization protocol (radar systems) are generic and free, but
{\em not} immediate.

Location systems are usually used either to allow a given observer
assigning  coordinates to particular events of his environment or to
allow every event of a given environment to know its proper
coordinates. Location systems constructed for the first of these two
functions, following their three-dimensional Newtonian analogues,
are called ({\em relativistic}) {\em reference systems}. In
relativity, where the velocity of transmission of information is
finite, they are necessarily not immediate. Poincar\'e-Einstein
location systems are reference systems in the present sense.

Location systems constructed for the second of these two functions
which, in addition, are generic, free and immediate, are called
\cite{coll-1}  ({\em relativistic}) {\em positioning systems.} Since
Poincar\'e-Einstein reference systems are the only known location
systems but they are not immediate, the first question is if in
relativity there exist positioning systems having the three demanded
properties of being generic, free and immediate. The
epistemic\footnote{The word {\em epistemic} is also used in the
sense of footnote 2.} answer is that there exists a little number of
them, their paradigmatic representative being constituted by four
clocks broadcasting their proper times\footnote{In fact, the
paradigmatic representative of positioning systems is constituted by
four point-like sources broadcasting countable electromagnetic
signals.}.

In Newtonian physics, when the velocity of transmission of
information is supposed infinite, both functions, of reference and
positioning, are {\em exchangeable} in the sense that data obtained
from any of the two systems may be transformed in data for the other
one. But this is no longer possible in relativity, where the
immediate character of positioning systems and the intrinsically
retarded  character of reference systems imposes a strong decreasing
hierarchy\footnote{In fact, whereas it is impossible to construct a
positioning system starting from a reference system (by transmission
of its data), it is always possible and very easy to construct a
reference system from a positioning system (it is sufficient that
every event sends its coordinate data to the observer.)}.

Consequently, in relativity the experimental or observational
context strongly conditions the function, conception and
construction of location systems. In addition, by their immediate
character, it results that whenever possible, there are positioning
systems, and not reference systems, which offer the most interest to
be constructed\footnote{For the Solar system, it has been recently
proposed a `galactic' positioning system, based on the signals of
four selected millisecond pulsars and a conventional origin. See
\cite{coll-tarantola}. For the neighborhood of the Earth, a primary,
auto-locating (see below in the text), fully relativistic,
positioning system has also been proposed, based on four-tuples of
satellited clocks broadcasting their proper time as well as the time
they receive from the others. The whole constellation of a global
navigation satellite system (GNSS), as union of four-tuples of
neighboring satellites, constitutes an atlas of local charts for the
neighborhood of the Earth, to which a global reference system
directly related to the conventional international celestial
reference system (ICRS) may be associated (SYPOR project). See
\cite{coll-1}.}.

What is the coordinate system physically realized by four clocks
broadcasting their proper time?  Every one of the four clocks
$\gamma_i$ ({\it emitters})  broadcasting his proper time $\tau^i,$
the future light cones of the points $\gamma_i(\tau^i)$ constitute
the coordinate hypersurfaces  $\tau^i=constant$ of the coordinate
system for some domain of the space-time. At every event of the
domain, four of these cones, broadcasting the times $\tau^i,$
intersect, endowing thus the event with the coordinate values
$\{\tau^i\}.$  In other words, the past light cone of every event
cuts the emitter world lines at $\gamma_i(\tau^i)$; then
$\{\tau^i\}$ are the {\it emission coordinates} of this event.

Let $\gamma$ be an observer equipped with a receiver allowing the
reception of the proper times $(\tau^i)$ at each point of his
trajectory. Then, this observer knows his trajectory in these
emission coordinates. We say then that this observer is a {\it
user} of the positioning system. It is worth pointing out that a
user could, eventually (but not necessary), carry a clock to
measure his proper time $\tau$.

A positioning system may be provided with the important quality of
being {\it auto-locating}. For this goal, the emitters have also to
be {\it transmitters} of the proper time they just receive, so that
at every instant they must broadcast their proper time {\em and}
also these other received proper times. Then, any user does not only
receive the emitted times $\{\tau^i\}$ but also the twelve
transmitted times $\{\tau^i_j\}$. These data allow the user to know
the trajectories of the emitter clocks in emission coordinates.

The interest, characteristics and good qualities of the relativistic
positioning systems have been  pointed out elsewhere \cite{coll-1,
coll-2, coll-3} and some explicit results have been recently
presented for the generic four-dimensional case \cite{coll-pozo,
pozo}. A full development of the theory for this generic case
requires a hard task and a previous training on simple and
particular examples. The two-dimensional approach that we present
here should help to better understand how these systems work as well
as the richness of the physical elements that this positioning
approach has.

Indeed, the two-dimensional case has the advantage of allowing the
use of precise and explicit diagrams which improve the qualitative
comprehension of the general four-dimensional positioning systems.
Moreover, two-dimensional constructions admit simple and explicit
analytic results.

It is worth remarking that the two-dimensional case has some
particularities and results that cannot be directly extended to the
generic four-dimensional case. Two dimensional constructions are
suitable for learning basic concepts about positioning systems, but
they do not allow to study some positioning features that
necessarily need a three-dimensional or a four-dimensional approach.

As already mentioned, the coordinate system that a positioning
system realizes is constituted by four null (one-parametric family
of) hypersurfaces, so that its covariant natural frame is
constituted by four null 1-forms. Such rather unusual {\em real null
frames} belong to the causal class $\{\, {\rm e}{\rm e}{\rm e}{\rm
e} \, , \, {\rm E}{\rm E}{\rm E}{\rm E}{\rm E}{\rm E} \, , \, \ell
\ell \ell \ell \, \}$ of frames among the 199 admitted by a
four-dimensional Lorentzian metric  \cite{coll-morales,
moralesERE05}. Up to recently, and all applications taken into
account, this causal class, or its algebraically dual one, $\{\,
{\rm l}{\rm l} {\rm l}{\rm l}\, , \,{\rm T}{\rm T}{\rm T}{\rm T}{\rm
T}{\rm T} \,  , \, e e e e \, \},$ has been considered in the
literature but very sparingly.

Zeeman \cite{Zeeman} seems the first to have used, for a technical
proof, real null frames, and Derrick found them as a particular case
of {\em symmetric frames} \cite{Derrick}, later extensively studied
by Coll and Morales \cite{coll-mor}. As above mentioned, they were
also those that proved that real null frames constitute a causal
class among the 199 possible ones.  Coll \cite{Yoluz} seems to have
been the first to propose the physical construction of coordinate
systems by means of light beams, obtaining real null frames as the
natural frames of such coordinate systems. Finkelstein and Gibbs
\cite{Finkelstein} proposed symmetric real null frames as a
checkerboard lattice for a quantum space-time. The physical
construction of relativistic coordinate systems `of GPS type', by
means of broadcasted light signals, with a real null coframe as
their natural coframe, seems also be first proposed by Coll
\cite{coll-2}. Bahder \cite{bahder} has obtained explicit
calculations for the vicinity of the Earth at first order in the
Schwarzschild space-time, and Rovelli \cite{Rovelli}, as
representative of a complete set of gauge invariant observables, has
developed the case where emitters define a symmetric frame in
Minkowski space-time. Blagojevi\`c et al. \cite{blago} analysed and
developed the symmetric frame considered in Finkelstein and Rovelli
papers.

All the last four references, as well as ours \cite{coll-1,
coll-tarantola, coll-3, coll-pozo, pozo, ferrandoERE05, cfm}, have
in common the awareness about the need of physically constructible
coordinate systems (location systems) in experimental projects
concerning relativity. But their future role, as well as their
degree of importance with respect to the up to now usual ones,
depends on the authors. For example, Bahder considers them as a way
to transmit to any user its coordinates with respect to an exterior,
previously given, coordinate system. Nevertheless, our analysis,
sketched above, on the generic, free and immediate properties of the
relativistic positioning systems lead us to think that they are
these systems which are assigned to become the {\em primary} systems
of any precision cartography. Undoubtedly, there is still a lot of
work to be made before we be able, as users, to verify and to
control this primary character, but the present general state of the
theory and the explicit results already known in two, three and
four-dimensional space-times encourage this point of view.  A key
concept for this primary character of a system, although not
sufficient, is the already mentioned of auto-location, whose
importance in the two-dimensional context is shown in the
propositions of this paper.

At first glance, relativistic positioning systems are nothing but
the relativistic model of the classical GPS (Global Positioning
System) but, as explained for example in \cite{coll-3}, this is not
so. In particular, the GPS uses its emitters (satellites) as simple
(and `unfortunately moving'!) beacons to transmit {\em another}
spatial coordinate system (the World Geodetic System 84) and an ad
hoc time scale (the GPS time), different from the proper time of the
embarked clocks, meanwhile for relativistic positioning systems the
unsynchronized proper times of the embarked clocks constitute the
fundamental ingredients of the system. As sketched in \cite{coll-1}
or \cite{coll-3}, positioning systems offer a new, paradigmatic, way
of decoupling and making independent the spatial segment of the GPS
system from its Earth control segment, allowing such a positioning
system to be considered as {\em the primary positioning reference}
for the Earth and its environment.

This work introduces in section II the basic elements of a
relativistic positioning system and lists the different kind of data
that the system generates and the users can obtain. In section III
it presents the explicit deduction of the emission coordinates from
a given null coordinate system where the proper time trajectories of
the emitters are known. Then, in section IV the positioning system
defined by two inertial emitters in flat space-time is studied, and
it is shown the simple but important result that the emitted and
transmitted times allow a complete determination of the metric in
emission coordinates. Section V is devoted to analyze the
information that a user of the relativistic positioning system can
obtain on the gravitational field, and it is shown that the emitted
data determine the gravitational field and its first derivatives
along the trajectories of the user and of the emitters. Finally, in
section VI the results are discussed and new problems, some open
ones and some others discussed elsewhere, are commented.

For the sake of completeness, in three appendices some basic results
about two-dimensional relativity in null coordinates are summarized.
A short communication of this work has been presented in the Spanish
Relativity meeting ERE-2005 \cite{ferrandoERE05}.

\section{Relativistic positioning systems: emission coordinates and
user data grid}

In a two-dimensional space-time, let $\gamma_1$ and $\gamma_2$ be
the world lines of two clocks measuring their proper times $\tau^1$
and $\tau^2$ respectively. Suppose that they broadcast them by means
of electromagnetic signals, and that the signals from each one of
the world lines reach the other.

\begin{figure}[htb]
    \includegraphics[angle=0,width=0.48\textwidth]{%
        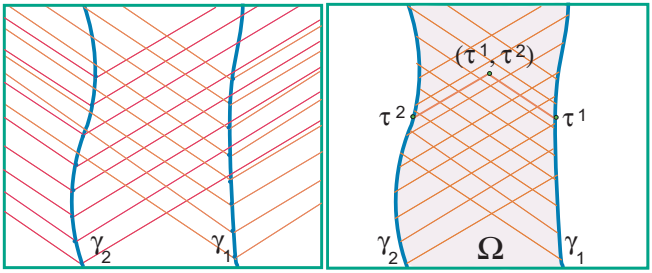}
    \caption{       \label{fig:light-cones}
        a) Every emitter $\gamma_i$ broadcasts
        his proper time $\tau^i$. b) Then, every event
        in the domain $\Omega$ between both emitters can be distinguished
        by the times $(\tau^1,\tau^2)$.}
\end{figure}

The signals carrying the times $\tau_1$ and $\tau_2,$ respectively
emitted at the events $\gamma_1(\tau^1)$ and $\gamma_2(\tau^2),$
that reach a third event cut at that event in the region
${\bar{\cal R}}$ between both clock world lines and they are
tangent if the event is outside this region. See Fig.
\ref{fig:light-cones}a.

According to the notions remembered and sketched in the
Introduction, such a system of two clocks ({\em emitters})
broadcasting their proper times constitutes a {\em relativistic
positioning system} in the two-dimensional space-time under
consideration.

The interior ${\cal R}$ of the above region ${\bar{\cal R}}$ may
be a domain (i.e. connected) or, if the clock world lines are
allowed to contact or to cut, a union of domains. Anyway, from now
on, we shall restrict our study to a sole domain, denoted by
$\Omega.$ Thus, according to the allowed situations, we may have
$\Omega \equiv {\cal R}$ or $\Omega \subset {\cal R}.$

The domain $\Omega$ constitutes a {\em coordinate domain}. Indeed,
every event  in it can be distinguished by the times $\tau^1$ and
$\tau^2$ received from the emitter clocks. See Fig.
\ref{fig:light-cones}b. In other words, the past light cone of
every event in $\Omega$ cuts the emitter world lines at
$\gamma_1(\tau^1)$ and $\gamma_2(\tau^2)$, respectively. Then
$(\tau^1,\tau^2)$ are the coordinates of the event. We shall refer
to them as the {\em emission coordinates} $\{\tau^1,\tau^2\}$ of
the system.
\begin{figure*}[htb]
    \includegraphics[angle=0,width=0.85\textwidth]{%
        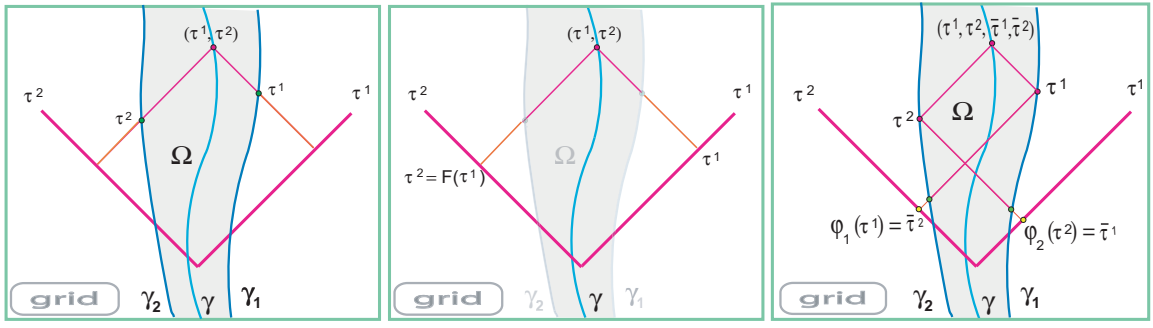}
    \caption{       \label{fig:grid-user}
            a) Geometric interpretation of the emission
            coordinates: in the grid $\{\tau^1\}\times \{\tau^2\}$
            the trajectory of a user $\gamma$ that receives the proper
            times $\{\tau^1, \tau^2\}$ broadcasted by the
            emitters $\gamma_1$, $\gamma_2$ is drawn. b) The {\em user positioning
            data} $\{\tau^1, \tau^2\}$ allow the user to know his trajectory
            in emission coordinates and he can draw it in the
            grid $\{\tau^1\}\times \{\tau^2\}$. c)  The {\em emitter positioning data}
            $\{\tau^1, \tau^2; \bar{\tau}^1, \bar{\tau}^2\}$ allow
        the user to know the emitter trajectories $\varphi_1(\tau^1)$,
        $\varphi_2(\tau^2)$ in emission coordinates $\{\tau^1, \tau^2\}$.}
\end{figure*}

On the contrary, all the events outside both world lines that
receive the same time $\tau^1$ receive also the same time
$\tau^2$  because both signals are parallel. See Fig.
\ref{fig:light-cones}a. Thus, these proper times do not
distinguish different events on  the segment of null geodesic
signals in the outside region: the signals $\tau^1$ and $\tau^2$
do not constitute coordinates for the events in the outside
region.

What happens {\em on} the clock world lines? The clock world lines
are (part of) the boundary of the domain $\Omega,$ so that they are
out of it. Nevertheless, there is no ambiguity to associate to their
instants, by continuity from $\Omega,$ a pair of proper times: the
one of the clock in question and that received from the other clock.
We shall refer to these pairs as the {\em emission coordinates of
the emitters}.

The coordinate lines (coordinate hypersurfaces) of the emission
coordinates $\{\tau^1,\tau^2\}$ are null geodesics. Consequently,
the emission coordinates are null coordinates. Thus, in emission
coordinates the space-time metric depends on the sole {\em metric
function} $m$ (see Appendix A):
\begin{equation} \label{metric-taus}
ds^2 = m(\tau^1,\tau^2) d \tau^1 d\tau^2
\end{equation}

An observer $\gamma,$ traveling throughout the emission coordinate
domain $\Omega$ and equipped with a receiver allowing the reading of
the received proper times $(\tau^1,\tau^2)$ at each point of his
trajectory, is called a {\it user} of the positioning system.

The essential physical components of the relativistic positioning
system are thus:
\begin{itemize}
\item[\rm(A)] Two emitters $\gamma_1$, $\gamma_2$ broadcasting their
proper times $\tau^1$, $\tau^2$.
\item[\rm(a)] Users $\gamma$ traveling in the domain $\Omega$
and receiving the broadcasted times $\{\tau^1, \tau^2\}.$
\end{itemize}

Observe that, as a {\em location system} (i.e. as a physical
realization of a mathematical coordinate system), the above
positioning system is {\em generic}, {\em free} and {\em
immediate} in the sense specified in the Introduction.

The plane $\{\tau^1\}\times\{\tau^2\}$ ($\tau^1, \tau^2 \in
\mathbb{R}$) in which the different data of the positioning system
can be transcribed is called the {\em grid} of the positioning
system. See Fig. \ref{fig:grid-user}a. Observe that not all the
points of the grid correspond to physical events. Only those limited
by the emitter trajectories can be physically detected.

It is worth remarking that any user receiving continuously the
emitted times $\{\tau^1, \tau^2\}$ knows his trajectory in the grid.
Indeed, from these {\em user positioning data} $\{\tau^1, \tau^2\}$
the user trajectory may be drawn in the grid (see Fig.
\ref{fig:grid-user}b), and its equation $F$,
\begin{equation}  \label{user-trajectory}
\tau^2 = F(\tau^1) \, ,
\end{equation}
may be extracted from these data.

Let us note that, whatever the user be, these data are
insufficient to construct both of the two emitter trajectories.

In order to give to any user the capability of knowing the emitter
trajectories in the grid, the positioning system must be endowed
with a device allowing every emitter to also broadcast the proper
time it is receiving from the other emitter. See Fig.
\ref{fig:grid-user}c. In other words, the clocks must be allowed to
broadcast {\em their emission coordinates}. As stated in the
Introduction, a positioning system so endowed is called an {\em
auto-locating positioning system}.

The physical components of an auto-locating positioning system are
thus:
\begin{itemize}
\item[(B)] Two emitters $\gamma_1$, $\gamma_2$
broadcasting their proper times $\tau^1,$ $\tau^2$ {\em and} the
proper times $\bar{\tau}^2$, $\bar{\tau}^1$ that they receive each
one from the other.
\item[(b)] Users $\gamma$, traveling in $\Omega$ and receiving these four
broadcasted times  $\{\tau^1, \tau^2; \bar{\tau}^1,
\bar{\tau}^2\}$.

\end{itemize}

Any user receiving continuously these {\em emitter positioning data}
$\{\tau^1, \tau^2; \bar{\tau}^1, \bar{\tau}^2\}$ may extract from
them not only the equation of   his trajectory in the grid, $\tau^2
= F(\tau^1),$ but also the equations of the trajectories of the
emitters (see Fig. \ref{fig:grid-user}c):
\begin{equation} \label{emitter-trajectories}
\varphi_1(\tau^1) = \bar{\tau}^2 \, , \qquad  \varphi_2(\tau^2) =
\bar{\tau}^1
\end{equation}

Eventually, the positioning system can be endowed with complementary
devices. For example, in obtaining the dynamic properties of the
system, it is necessary to know the acceleration of the emitters.
For this ability:
\begin{itemize}
\item[(C)] The emitters $\gamma_1$, $\gamma_2$
carry accelerometers and broadcast their acceleration $\alpha_1$,
$\alpha_2$.
\item[(c)] The users $\gamma$, in addition to the emitter positioning data,
also receive the emitter accelerations $\{\alpha_1, \alpha_2\}$.
\end{itemize}
These new elements allow any user (and, in particular, the
emitters) to know the acceleration scalar of the emitters:
\begin{equation} \label{emitter-accelerations}
\alpha_1 = \alpha_1(\tau^1) \, , \qquad \alpha_2 = \alpha_2(\tau^2)
\end{equation}
\begin{figure*}[bht]
    \includegraphics[angle=0,width=0.85\textwidth]{%
        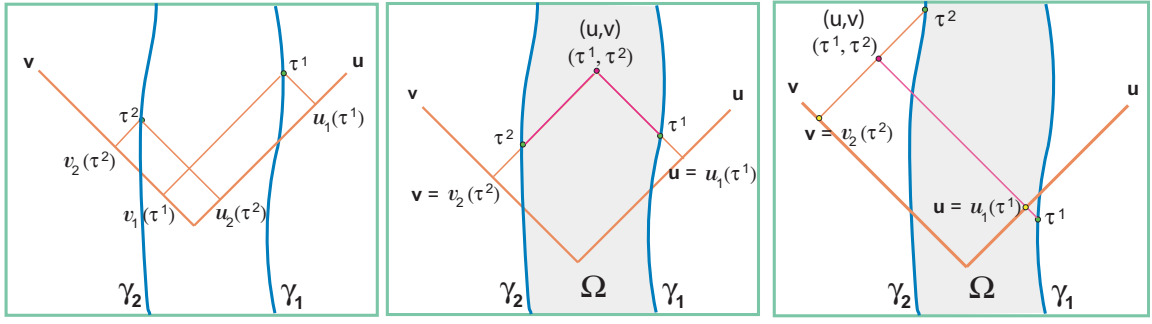}
    \caption{       \label{fig:emitters-uv}
        a) Proper time history of two emitters $\gamma_1$, $\gamma_2$ in
        a null system $\{\texttt{u},\texttt{v}\}$. b) Geometric interpretation
        of the coordinate change between emission coordinates $\{\tau^1,\tau^2\}$
        and a null system $\{\texttt{u},\texttt{v}\}$. c) Outside the domain
        $\Omega$, $\{\tau^1,\tau^2\}$ are still null coordinates but not more emission coordinates.}
\end{figure*}

In some cases, it can be useful that the users generate their own
data:
\begin{itemize}
\item[(d)] Any user carries a clock that measures his proper time $\tau$.
\item[(e)] Any user carries an accelerometer that measures his acceleration $\alpha$.
\end{itemize}
The user's clock allows any user to know his proper time function
$\tau(\tau^1)$ (or $\tau(\tau^2)$) and, consequently by using
(\ref{user-trajectory}), to obtain the proper time parametrization
of his trajectory:
\begin{equation}
\gamma \equiv \begin{cases} \  \tau^1 = \tau^1(\tau) \\ \ \tau^2 =
\tau^2(\tau) \end{cases}
\end{equation}
The user's accelerometer allows any user to know his proper
acceleration scalar:
\[
\alpha = \alpha(\tau)
\]

Thus, a relativistic positioning system may be performed in such a
way that any user can obtain  a subset of the {\it user data}:
\begin{equation} \label{user-data}
\{\tau^1, \tau^2; \bar{\tau}^1, \bar{\tau}^2; \alpha_1, \alpha_2;
\tau, \alpha\}
\end{equation}

It is worth remarking that the pairs of data $\{\tau^1,
\bar{\tau}^2\}$ and $\{\tau^2, \bar{\tau}^1\}$ which give the
emitter trajectories (\ref{emitter-trajectories}) do not depend on
the user that receives them, i.e., every user draws in his user grid
the same emitter trajectories. A similar think occurs with the pairs
$\{\tau^1, \alpha_1\}$ and $\{\tau^2, \alpha_2\}$ which give the
emitter accelerations (\ref{emitter-accelerations}). Thus, among the
{\em user data} (\ref{user-data}) we can distinguish the subsets:
\begin{itemize}
\item
{\em emitter positioning data} $\{\tau^1, \tau^2; \bar{\tau}^1,
\bar{\tau}^2\}$,
\item
{\it public data} $\{\tau^1, \tau^2; \bar{\tau}^1, \bar{\tau}^2;
\alpha_1, \alpha_2\}$,
\item
{\it user proper data} $\{\tau,
\alpha\}$.
\end{itemize}

The grid of the positioning system plays an important role in
practical positioning. Auto-locating systems allow any user to
determine the domain $\Omega$ in the grid where the parameters
$\{\tau^1, \tau^2\}$ constitute effectively a emission coordinate
system, i.e. may be physically detected.

It is to be noted that in a positioning system the emitter clocks
are constrained to measure their proper time, but otherwise their
time scale (their origin) is completely independent one of the
other, that is to say, they are not submitted to any prescribed {\em
synchronization.} The {\em form} of the trajectories of emitters and
users in the grid is an {\em invariant} of the independently chosen
time scales, only their {\em position} in the grid translates with
them.

\section{Construction of the emission coordinates}

In a generic two-dimensional space-time let
$\{\texttt{u},\texttt{v}\}$ be an arbitrary null coordinate
system, i.e., a coordinate system where the metric interval can be
written (see appendix \ref{ap-A}):
\[d s^2 = \texttt{m}(\texttt{u}, \texttt{v})\, d\texttt{u} \, d\texttt{v} \]

Let us assume the {\it proper time history of two emitters} to be
known in this coordinate system (see Fig. \ref{fig:emitters-uv}a):
\begin{equation}  \label{principalemit}
\gamma_1 \equiv \begin{cases} \texttt{u} = u_1(\tau^1) \\
\texttt{v} = v_1(\tau^1)\end{cases} \qquad \gamma_2 \equiv
\begin{cases} \texttt{u} = u_2(\tau^2) \\ \texttt{v} = v_2(\tau^2)
\end{cases}
\end{equation}

We can introduce the proper times as coordinates
$\{\tau^1,\tau^2\}$ by defining the change to the null system
$\{\texttt{u},\texttt{v}\}$ given by (see Fig.
\ref{fig:emitters-uv}b):
\begin{equation}  \label{coordinatechange0}
\begin{array}{l}
\texttt{u} = u_1(\tau^1) \\
\texttt{v} = v_2(\tau^2)
\end{array}
\qquad \quad
\begin{array}{l}
\tau^1 = u_1^{-1}(\texttt{u}) = \tau^1(\texttt{u}) \\
\tau^2 = v_2^{-1}(\texttt{v}) = \tau^2(\texttt{v})
\end{array}
\end{equation}

In terms of the coordinates $\{\texttt{u},\texttt{v}\}$, the
region $\Omega$ where the new coordinates $\{\tau^1,\tau^2\}$ can
be considered emitted times is (see Fig. \ref{fig:emitters-uv}b):
\[
\Omega \equiv  \big\{ (\texttt{u}, \texttt{v}) \ |   \
F_2^{-1}(\texttt{v}) \leq \texttt{u} \, , \quad  F_1(\texttt{u})
\leq \texttt{v} \big\}
\]
where $F_i$ are the {\it emitter trajectory functions}:
\begin{equation}  \label{principaltraj}
\texttt{v} = F_i(\texttt{u}) \, ,  \qquad F_i = v_i \circ u_i^{-1}
\end{equation}

Thus, relations (\ref{coordinatechange0}) define {\it emission
coordinates} in the {\it emission coordinate domain} $\Omega$.

In the region outside $\Omega$ the change (\ref{coordinatechange0})
also determines null coordinates which are an extension of the
emission coordinates. But in this region the coordinates are not
physical, i.e. are not the emitted proper times of the emitters
$\gamma_1$, $\gamma_2$ . See Fig. \ref{fig:emitters-uv}c. Note that
there are several null coordinate systems that can be associated
with a unique observer or with two observers by considering their
advanced or retarded signals. But here we are limited to emission
coordinates: those generated by a positioning system and thus by
retarded signals.

\begin{figure}[hbt]
    \includegraphics[angle=0,width=0.48\textwidth]{%
        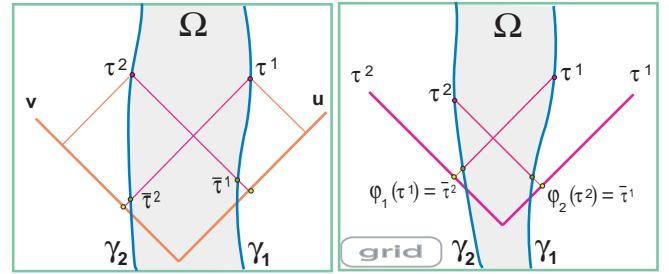}
    \caption{       \label{fig:emitters-uv-taus}
        Emitter trajectories in a coordinate diagram $\{\texttt{u},\texttt{v}\}$
        and in the grid $\{\tau^1,\tau^2\}$.}
\end{figure}

In emission coordinates, the emitter trajectories take the
expression (see Fig. \ref{fig:emitters-uv-taus}):
\begin{equation}  \label{principalemitintaus0}
\gamma_1 \equiv \begin{cases} \tau^1 = \tau^1 \\ \tau^2 =
\varphi_1(\tau^1)\end{cases} \qquad \gamma_2 \equiv
\begin{cases} \tau^1 = \varphi_2(\tau^2) \\ \tau^2 = \tau^2
\end{cases}
\end{equation}
where, from (\ref{principalemit}) and (\ref{coordinatechange0}), the
functions $\varphi_i$ giving the emitter trajectories are given by:
\begin{equation}  \label{phis}
\varphi_1 = v_2^{-1} \circ v_1    \, , \qquad   \varphi_2 = u_1^{-1}
\circ u_2
\end{equation}

The {\it principal observers} associated with an arbitrary null
system are those whose proper time coincides with one of the two
null coordinates (see appendix \ref{ap-C}). The expression
(\ref{principalemitintaus0}) shows that:
\begin{result}
The two emitters are particular principal observers of the emission
coordinate system that they define.
\end{result}

Finally, if we know the metric function $\texttt{m}(\texttt{u},
\texttt{v})$ in null coordinates $\{\texttt{u},\texttt{v}\}$, we can
obtain the metric interval in emission coordinates
$\{\tau^1,\tau^2\}$ by using the change (\ref{coordinatechange0}):
\begin{equation} \label{metric-change}
\begin{array}{c}
ds^2 = m(\tau^1,\tau^2) d \tau^1 d\tau^2 \\[4mm]
m(\tau^1,\tau^2) = \texttt{m}(u_1(\tau^1),  v_2(\tau^2))
u_1'(\tau^1) v_2'(\tau^2)
\end{array}
\end{equation}

\begin{figure*}[tbh]
    \includegraphics[angle=0,width=0.85\textwidth]{%
        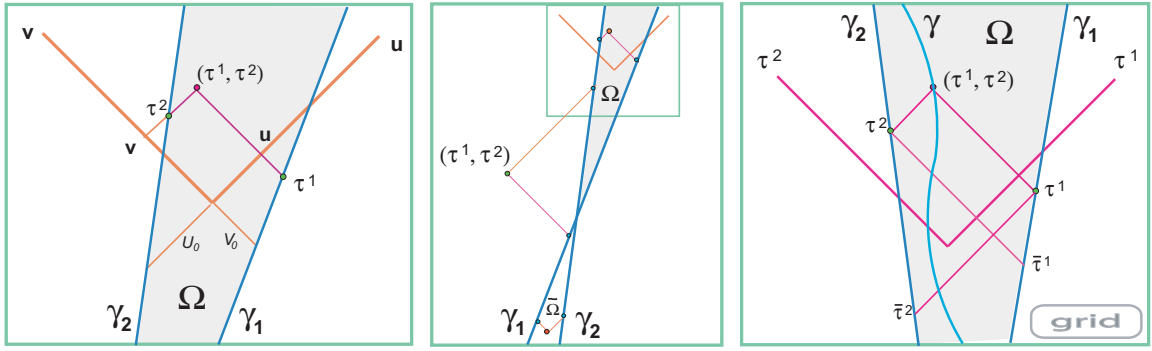}
    \caption{       \label{fig:geodesic-emitters}
        a) Proper time parametrization of geodesic
        emitters  $\gamma_1$, $\gamma_2$ and associated emission coordinates $\{\tau^1,\tau^2\}$. b)
        The coordinates $\{\tau^1,\tau^2\}$ defined in (\ref{coordinatechangeinertial}) can be
        extended, but outside of $\Omega$ they are not emission coordinates. c) Trajectories of geodesic
        emitters in the grid, where the `initial' inertial information does not appear.}
\end{figure*}

\section{Positioning with inertial emitters}

In this section we consider the simple example of a positioning
system defined by {\it two inertial emitters} $\gamma_1$,
$\gamma_2$ in {\it flat space-time}.

In inertial null coordinates $\{\texttt{u},\texttt{v}\}$ the
trajectory of the emitters are (see Fig.
\ref{fig:geodesic-emitters}a):
\[
\begin{array}{lcl}
 \gamma_1 & : &  \qquad  \texttt{v} =
\displaystyle \frac{1}{\lambda_1^2}
\texttt{u} + v_0\\[3mm]
\gamma_2 & : & \qquad   \texttt{v} = \displaystyle
\frac{1}{\lambda_2^2}( \texttt{u} - u_0)
\end{array}
\]
where $\lambda_1$, $\lambda_2$ are the shifts of the emitters with
respect to the inertial system. We could choose these inertial
coordinates so that one emitter be at rest. The constants $u_0$ and
$v_0$ can also be arbitrarily chosen depending on the inertial
origin. But, for the moment, we will consider that they take
arbitrary values. In Fig. \ref{fig:geodesic-emitters} we have take
$u_0 <0$, $v_0<0$ and $1 \leq \lambda_2 \leq \lambda_1$.

The origin of the emitter proper times can be taken such that the
proper time history of the emitters be:
\begin{equation}  \label{inertialprincipalemit}
\gamma_1 \equiv \begin{cases} \, \texttt{u} = \lambda_1 \tau^1
\\[1mm]
\, \texttt{v} = \displaystyle \frac{1}{\lambda_1} \tau^1 +
v_0\end{cases} \  \gamma_2 \equiv \begin{cases} \, \texttt{u} =
\lambda_2 \tau^2 +
u_0 \\[1mm] \, \texttt{v} =\displaystyle  \frac{1}{\lambda_2} \tau^2 \end{cases}
\end{equation}
Then, taking into account the construction presented in previous
section, the coordinate transformation from the inertial coordinate
system $\{\texttt{u},\texttt{v}\}$ to the emission coordinate system
$\{\tau^1,\tau^2\}$ is given by (see Fig.
\ref{fig:geodesic-emitters}a):
\begin{equation}  \label{coordinatechangeinertial}
\begin{array}{lll}
\texttt{u}  = u_1(\tau^1) = \lambda_1 \tau^1 & \qquad \qquad & \tau^1 =  \displaystyle \frac{1}{\lambda_1} \, \texttt{u}  \\[2mm]
\texttt{v}  = v_2(\tau^2) = \displaystyle  \frac{1}{\lambda_2}
\tau^2 & \qquad \qquad & \tau^2 =  \lambda_2\texttt{v}
\end{array}
\end{equation}

From this transformation, the metric tensor in emission
coordinates $\{\tau^1,\tau^2\}$ can be obtained. Indeed, from
(\ref{metric-change}) we have:
\[
m(\tau^1,\tau^2) = u_1'(\tau^1)v_2'(\tau^2) =
\frac{\lambda_1}{\lambda_2}
\]
Thus, the metric function is constant and equals the relative shift
between both emitters:
\[
ds^2 = \lambda \, d \tau^1 d \tau^2 \, , \qquad \lambda \equiv
\frac{\lambda_1}{\lambda_2}
\]

It is worth pointing out that the coordinates $\{\tau^1,\tau^2\}$
defined in (\ref{coordinatechangeinertial}) can be extended
throughout the whole space-time (see Fig.
\ref{fig:geodesic-emitters}b), but that outside the domain $\Omega$
they are not emission coordinates. Nor are they in the domain
$\tilde{\Omega}$ in Fig. \ref{fig:geodesic-emitters}b, where they
are ``advanced-advanced'' coordinates.

Of course, in this domain $\tilde{\Omega}$ the emitters $\gamma_1$,
$\gamma_2$ also define emission coordinates, but they are not given
by (\ref{coordinatechangeinertial}). In this case we must
interchange the role of both emitters. Then, in $\tilde{\Omega}$ the
emission coordinates $\{\tilde{\tau}^1,\tilde{\tau}^2\}$ are given
by:
\begin{equation}  \label{coordinatechangeinertialbis}
\begin{array}{l}
\texttt{u}  = \lambda_2 \tilde{\tau}^1 + u_0 \\[1mm]
\texttt{v}  =\displaystyle  \frac{1}{\lambda_1} \tilde{\tau}^2 + v_0
\end{array}
\end{equation}
and the metric tensor is now:
\[
ds^2 = \frac{1}{\lambda} \, d \tilde{\tau}^1 d \tilde{\tau}^2 \, ,
\qquad \lambda \equiv \frac{\lambda_1}{\lambda_2}
\]

The coordinates $\{\tilde{\tau}^1,\tilde{\tau}^2\}$ can also be
extended to the whole space-time, but only on the domain
$\tilde{\Omega}$ are emission coordinates.

Note that the extensions of the emission coordinates
(\ref{coordinatechangeinertial}) and
(\ref{coordinatechangeinertialbis}) are different everywhere when
$\lambda \not=1$. When the emitters are at rest with each other
($\lambda =1$), both extensions coincide (up to an origin change)
and they define inertial coordinates.

Let us return now to our domain $\Omega$ where the emission
coordinates are given by (\ref{coordinatechangeinertial}). In these
coordinates $\{\tau^1,\tau^2\}$, the emitter trajectories are (see
Fig. \ref{fig:geodesic-emitters}c):
\begin{eqnarray}  \label{emit-taus-inertial-1}
\gamma_1 \equiv \begin{cases} \tau^1 = \tau^1 \\ \tau^2 =
\varphi_1(\tau^1) \equiv \displaystyle  \frac{1}{\lambda} \tau^1 +
\tau^2_0
\end{cases}
\\[1mm]  \label{emit-taus-inertial-2}
\gamma_2 \equiv \begin{cases} \tau^1 = \varphi_2(\tau^2) \equiv
\displaystyle \frac{1}{\lambda} \tau^2 + \tau^1_0 \\ \tau^2 = \tau^2
\end{cases}
\end{eqnarray}

Observe that these emitter trajectories in the grid have lost the
`initial' inertial information of the null coordinate system
$\{\texttt{u},\texttt{v}\}$. Indeed, only the shift $\lambda$
between the emitters appears (and not the relative shift $\lambda_i$
of each emitter with respect to the inertial system), and the grid
origin depends exclusively on the choice of origin of the emitter
times (not on the inertial coordinate origin). See Fig.
\ref{fig:geodesic-emitters}c for the case $\lambda \not= 0$; for
emitters at rest with each other the emitter trajectories become
parallel.

What information can a user obtain from the public data? Evidently
$\{\tau^1, \tau^2\}$ place the user on the grid, and $\{\tau^1,
\tau^2, \bar{\tau}^1, \bar{\tau}^2\}$, $\bar{\tau}^i =
\varphi_j(\tau^j)$, make the same for the emitters.

On the other hand, we can observe in (\ref{emit-taus-inertial-1}),
(\ref{emit-taus-inertial-2}) that the metric function $m(\tau^1,
\tau^2) = \lambda$ can be obtained from the emitter positioning data
$\{\tau^1_P, \tau^2_P; \bar{\tau}^1_P, \bar{\tau}^2_P\}$ and
$\{\tau^1_Q, \tau^2_Q; \bar{\tau}^1_Q, \bar{\tau}^2_Q\}$ at two
events $P$ and $Q$. Thus we have:
\begin{proposition}
In terms of the emitter positioning data $\{\tau^1, \tau^2;
\bar{\tau}^1, \bar{\tau}^2\}$, the space-time metric is given by
\begin{equation}
\label{general} ds^2 = \sqrt{\frac{\Delta\tau^1
\Delta\tau^2}{\Delta\bar{\tau}^1 \Delta\bar{\tau}^2}} \, d \tau^1 d
\tau^2
\end{equation}
where $\Delta\tau^i,$ $\Delta\bar{\tau}^i$ stand for the differences
of values at two events.
\end{proposition}

Although simple, this statement is very important because it shows
that, if the emitters are geodesic, the sole public quantities
$\{\tau^1, \tau^2; \bar{\tau}^1, \bar{\tau}^2\}$ received by any
user allow him to know the space-time metric everywhere.

Observe that when both emitter clocks are synchronized such that
they indicate the instant zero at the virtual cut event\footnote{Let
us note that the cut event does not belong to the domain $\Omega$ of
the positioning data. It is obtained by {\em virtual} prolongation
of the emitter positioning data.} ($\tau^1_c = \tau^2_c=0$), the
user only needs to know the emitter positioning data at one event
and the space-time metric reduces then to\footnote{This expression
is due to A. Tarantola (private communication).}:
\begin{equation}
\label{menosgeneral} ds^2 = \sqrt{\frac{\tau^1 \tau^2}{\bar{\tau}^1
\bar{\tau}^2}}  \, d \tau^1 d \tau^2
\end{equation}

Note that even if we know that the clocks are geodesic,
(\ref{menosgeneral}) cannot be applied unless the clocks be
synchronized in the way above mentioned {\em and} the users know
this fact, meanwhile (\ref{general}) is valid whatever be the
synchronization of the geodesic clocks. Furthermore, the
trajectories of the clocks in the grid being then known and the
metric being given by (\ref{general}), a simple computation allows
to know the precise synchronization of the clocks (i.e. the values
$\tau^1_c$ and $\tau^2_c$ of the virtual cut event).

The geodesic character of the clocks may be an a priori information
forming part of the dynamical characteristics of the clocks and the
foreseen control of the system, or may be also a real time
information  if clocks and users are endowed with devices allowing
the users to know the emitter accelerations $\{\alpha_1,
\alpha_2\}$. In any case, the user information of the geodesic
character of the clocks by any of these two methods is generically
necessary, because {\em emitter trajectories that are straight lines
in the user grid are not necessarily geodesic trajectories in the
space-time} \cite{cfm}.

\section{Gravimetry and positioning}

We already know that, whatever be the curvature of the
two-dimensional space-time, the {\it emitter positioning data}
\begin{equation} \label{emit-pos-data}
\{\tau^1, \tau^2; \bar{\tau}^1, \bar{\tau}^2\}
\end{equation}
determine, on one hand, the {\it user trajectory}
\[
\tau^2 = F(\tau^1)
\]
and, on the other hand, the {\it emitter trajectories} $\varphi_i$,
\[
\bar{\tau}^2 = \varphi_1(\tau^1) \, , \qquad \bar{\tau}^1 =
\varphi_2(\tau^2)
\]

But, what about the metric interval when the user has no information
about the gravitational field? In other words, what gravimetry can a
user do from the data offered by a positioning system?

From the sole emitter positioning data (\ref{emit-pos-data}) and
denoting by a dot the derivative with respect to the corresponding
proper time, one has:
\begin{proposition} \label{prop-emes}
The emitter positioning data $\{\tau^1, \tau^2; \bar{\tau}^1,
\bar{\tau}^2\}$ determine the space-time metric function along the
emitter trajectories, namely:
\begin{eqnarray} \label{metrictrajectories1}
m(\tau^1, \varphi_1(\tau^1)) = \frac{1}{\dot{\varphi}_1(\tau^1)}
\\
m(\varphi_2(\tau^2) , \tau^2) = \frac{1}{\dot{\varphi}_2(\tau^2)}
\label{metrictrajectories2}
\end{eqnarray}
\end{proposition}

This result follows from the fact that emitters are also principal
observers for the emission coordinates and the kinematic expression
(\ref{C-unit}) for these observers.

If in addition of the emitter positioning data (\ref{emit-pos-data})
the user knows the emitter accelerations $\{\alpha_1, \alpha_2\}$,
and thus the acceleration scalars $\alpha_1(\tau^1)$,
$\alpha_2(\tau^2)$, one has also the following metric information:
\begin{proposition} \label{prop-grad}
The public data $\{\tau^1, \tau^2; \bar{\tau}^1, \bar{\tau}^2;
\alpha_1, \alpha_2\}$ determine the gradient of the space-time
metric function along the emitter trajectories, namely:
\begin{equation}  \label{metricacceleration1}
\hspace{-1.3mm}
\begin{array}{l}
(\ln m),_1(\tau^1, \varphi_1(\tau^1))  =  \alpha_1(\tau^1) \\[2mm]
\displaystyle (\ln m),_2(\tau^1, \varphi_1(\tau^1))  =
\frac{-1}{\dot{\varphi}_1(\tau^1)} \left[\alpha_1(\tau^1) +
\frac{\ddot{\varphi}_1(\tau^1)}{\dot{\varphi}_1(\tau^1)}\right]
\end{array}
\end{equation}
\begin{equation}  \label{metricacceleration2}
\hspace{-1.3mm}
\begin{array}{l}
\displaystyle (\ln m),_1(\tau^2, \varphi_2(\tau^2))  =
\frac{1}{\dot{\varphi}_2(\tau^2)} \left[\alpha_2(\tau^2) -
\frac{\ddot{\varphi}_2(\tau^2)}{\dot{\varphi}_2(\tau^2)}\right] \\[6mm]
(\ln m),_2(\varphi_2(\tau^2) , \tau^2) = -\alpha_2(\tau^2)
\end{array}
\end{equation}
\end{proposition}

This result follows from  expression (\ref{acc-scalar-po}) of the
scalar acceleration of the principal observers and the fact that
once the metric function $m(\tau^1, \tau^2)$ is known on a
trajectory, the knowledge of its gradient needs only of one
transverse partial derivative.

Alternatively, even when the positioning system is not
auto-locating, i.e., it broadcast only the proper times $\{\tau^1,
\tau^2\}$, if in addition the user also knows the proper user data
$\{\tau, \alpha\}$, he may obtain gravimetric information along his
trajectory. From his proper time $\tau$ measured by his clock, he
can know his proper time function, say $\tau = \tau(\tau^1)$, by
comparing $\tau$ with the proper time $\tau^1$ received from the
emitter $\gamma_1$; consequently he can obtain his parameterized
proper time trajectory:
\[
\begin{array}{l}
\tau^1 = \tau^1(\tau) \\
\tau^2 = \tau^2(\tau)
\end{array}
\]
From his accelerometer he can obtain his acceleration scalar
$\alpha(\tau)$. Then, the user may have the following gravimetry
information:
\begin{proposition}
The public-user data $\{\tau^1, \tau^2; \tau, \alpha\}$ determine
the space-time metric function and its gradient on the user
trajectory, namely:
\begin{equation} \label{metricuser}
m(\tau^1(\tau), \tau^2(\tau))  = \frac{1}{\dot{\tau}^1(\tau)
\dot{\tau}^2(\tau)}
\end{equation}
\begin{equation} \label{metricaccelerationuser}
\begin{array}{l}
\displaystyle (\ln m),_1(\tau^1(\tau), \tau^2(\tau)) =
\frac{1}{\dot{\tau}^1(\tau)}\left[\alpha(\tau) -
\frac{\ddot{\tau}^1(\tau)}{\dot{\tau}^1(\tau)}\right]\\[3mm]
\displaystyle (\ln m),_2(\tau^1(\tau), \tau^2(\tau))  = -
\frac{1}{\dot{\tau}^2(\tau)}\left[\alpha(\tau) +
\frac{\ddot{\tau}^2(\tau)}{\dot{\tau}^2(\tau)}\right]
\end{array}
\end{equation}
\end{proposition}

This result follows from expressions (\ref{B-unit}) and
(\ref{B-ac-scalar}) on kinematics in null coordinates.

As an example of these kinds of information, let us consider a user,
with no previous information on the gravitational field, and
receiving the following specific public data $\{\tau^1, \tau^2;
\bar{\tau}^1, \bar{\tau}^2; \alpha_1, \alpha_2\}$, namely emitter
positioning data $\{\tau^1, \tau^2; \bar{\tau}^1, \bar{\tau}^2\}$
showing a particular linear relation between the $\bar{\tau}$'s and
the $\tau$'s:
\begin{equation} \label{grav-slope}
\begin{array}{l}
\bar{\tau}^2 = \varphi_1(\tau^1) \equiv  \displaystyle
\frac{1}{\lambda} \tau^1 +
\tau^2_0\\ \vspace{-2mm} \\
\bar{\tau}^1 = \varphi_2(\tau^2) \equiv \displaystyle
\frac{1}{\lambda} \tau^2 + \tau^1_0
\end{array}
\end{equation}
with $\lambda > 1$ and vanishing public data $\{\alpha_1,
\alpha_2\}$:
\begin{equation} \label{grav-slope-b}
\alpha_1(\tau^1) = 0 \, ,  \quad  \alpha_2(\tau^2) = 0
\end{equation}
Then, proposition \ref{prop-emes} gives the following metric
function along the emitter trajectories:
\[m(\tau^1) = \lambda \ , \qquad m(\tau^2) = \lambda  \, ,\]
and proposition \ref{prop-grad} implies that the gradient of the
metric function vanishes  along the emitter trajectories:
\[
m,_1(\tau^1) = m,_2(\tau^1) = 0 \, , \qquad  m,_1(\tau^2) =
m,_2(\tau^2) = 0
\]

Let us note that the above specific public data (\ref{grav-slope}),
because of their complementary slopes, coincide with those generated
by the positioning system defined by two geodesic emitters in flat
space-time, as considered in section IV (see expressions
(\ref{emit-taus-inertial-1}), (\ref{emit-taus-inertial-2}), and Fig.
\ref{fig:geodesic-emitters}c). Nevertheless, it is to be remarked
that the same specific emitter positioning data $\{\tau^1, \tau^2;
\bar{\tau}^1, \bar{\tau}^2\}$ verifying (\ref{grav-slope}) could be
obtained from non geodesic emitters in flat space-time as well as by
geodesic or not geodesic emitters in a non flat space-time. Also,
the same specific public data $\{\tau^1, \tau^2; \bar{\tau}^1,
\bar{\tau}^2; \alpha_1, \alpha_2\}$ verifying (\ref{grav-slope}) and
(\ref{grav-slope-b}) could be obtained from geodesic emitters in a
non flat space-time. This point will be analyzed in detail elsewhere
\cite{cfm}.

\section{Discussion and work in progress \label{discussion}}

In this work we have explained the basic features of relativistic
positioning systems in a two-dimensional space-time (section II) and
have obtained the analytic expression of the emission coordinates
associated with such a system (section III). In order to a better
understanding of these systems we have developed an example in
detail: the positioning system defined in flat space-time by
geodesic emitters, showing the striking result that the emitted and
transmitted times allow a complete determination of the metric in
emission coordinates (section IV). Finally, we have shown that the
data that a user obtains from the positioning system in arbitrary
space-times determine the gravitational field and its gradient along
the emitters and user world lines (section V).

It is worth remarking that the extension to the four-dimensional
case of the two-dimensional methods used here needs of additional
information. In particular, the information of the angles between
pairs of the arrival signals broadcasted by the clocks could help in
the obtention of the gravitational field along the trajectory of the
user \cite{coll-pozo, pozo}.

Nevertheless, the two-dimensional approach presented here helps
strongly to the understanding of how positioning systems work and
what is the physical role of their basic elements. We have used
explicit diagrams that improve the qualitative comprehension of
these systems and we have obtained simple analytic results. These
advantages encourage in putting and solving new two-dimensional
problems, many of them appearing in a natural way from the results
presented in this work.

Elsewhere \cite{cfm} we consider positioning systems in flat metric
other than that defined by geodesic emitters. Our first results on
this matter show interesting behaviors. For example, trajectories of
uniformly accelerated emitters are parallel straight lines in the
grid, as also happens with the trajectories of some other emitters
with more complicated acceleration laws.

Moreover, as a first contact to understand the behavior of
relativistic positioning systems in a non flat gravitational field,
we study positioning systems defined in the Schwarzschild plane by
two stationary emitters \cite{cfm}.

The study of these two cases of uniformly accelerated emitters
brings to light an interesting situation: the emitter positioning
data of both systems lead to an identical public grid. How a user
can distinguish both systems? In \cite{cfm} we analyze this question
and show that the full set of the user data determine the
Schwarzschild mass. This simple two-dimensional example suggests
that the relativistic positioning systems could be useful in
gravimetry for reasonable parameterized models of the gravitational
field.

This gravimetry case is only a particular situation of the general
gravimetry problem in Relativity where user data are the unique
information that a user has. We have shown here that these data
allow the user to obtain the metric function and its gradient on
some trajectories. This information on the gravitational field can
be increased introducing `secondary' emitters, that broadcast the
information they receive from the system, allowing any user to know
the metric function and its gradient on these additional
trajectories.

Some circumstances can lead us to take the point of view where the
user knows the space-time in which he is immersed (Minkowski,
Schwarzschild,...) and he wants to obtain, from the user data,
information on his local unities of time and distance, his
acceleration, the metric expression in the emission coordinates, and
his trajectory and emitters trajectories in a characteristic
coordinate system of the given space-time (inertial in flat metric,
stationary coordinates in Schwarzschild o other stationary
metric,...).

We undertake this problem for the flat case in \cite{cfm}, where we
analyze the minimum set of data that determine all this system and
user information. A striking result is that the user data are not
independent quantities: if we know the emitter positioning data,
then the accelerations of the emitters and of the user along their
trajectories are determined by the sole knowledge of one
acceleration during only an echo-causal interval between the emitter
trajectories.

\begin{acknowledgments}
This work has been supported by the Spanish Ministerio de
Educaci\'on y Ciencia, MEC-FEDER project AYA2003-08739-C02-02.
\end{acknowledgments}

\appendix

\section{Two-dimensional metrics in null coordinates}
\label{ap-A}

Here several simple and general results about two-dimensional
metrics are summarized and some usual relativistic questions are
developed by using the so called null coordinates.

In a flat two-dimensional space-time, with every inertial coordinate
system $\{t,x\}$ we can associate {\it inertial null coordinates}
$\{\texttt{u},\texttt{v}\}$:
\[\begin{array}{ll}
\texttt{u} = t + x       \  & \qquad \qquad  t = \frac12
(\texttt{u} + \texttt{v}) \
\\[1mm]
\texttt{v} = t - x   \ &  \qquad \qquad  x = \frac12 (\texttt{u} -
\texttt{v}) \
\end{array}\]
In coordinates $\{\texttt{u},\texttt{v}\}$, the metric tensor
takes the form:
\[d s^2 = dt^2 - d x^2 = d\texttt{u} \, d\texttt{v}\]

A boost between two inertial systems $\{t,x\}$,
$\{\bar{t},\bar{x}\}$ with a relative velocity $\beta=\tanh \psi$
takes a simple expression in inertial null coordinates
$\{\texttt{u},\texttt{v}\}$,
$\{\bar{\texttt{u}},\bar{\texttt{v}\}}$:
\begin{equation}
\begin{array}{l}  \label{nullboost}
\bar{\texttt{u}} = e^{\psi} \texttt{u}  \\[1mm]
\bar{\texttt{v}} = e^{-\psi} \texttt{v}
\end{array}
\end{equation}
Let us note that the factor
\[s=e^{\psi} = \sqrt{\frac{1+ \beta}{1-\beta}} = 1+z\]
is the {\it shift parameter} between both inertial systems.

An important result in two-dimensional Riemannian geometry states
that {\it every two-dimensional metric is (locally) conformally
flat}. In the Lorentzian case, starting from the inertial systems
associated with the flat metric conformal to the given space-time
metric, we see that {\it null coordinates}
$\{\texttt{u},\texttt{v}\}$ exist such that
\[d s^2 = m(\texttt{u}, \texttt{v})\, d\texttt{u} \, d\texttt{v} \]

In a two-dimensional geometry $g$ the scalar curvature $R(g)$ is
the unique strict component of the Riemann tensor. In terms of the
conformal factor $m$ the scalar curvature is:
\[R(g) = \frac1m \Delta \ln m = \Delta_g \ln m \, , \qquad  g=m
\eta\]
where $\Delta$ is the Laplacian operator for the flat metric
$\eta$. From here, it follows that {\it a two-dimensional metric
is flat iff the logarithm of the conformal factor is an harmonic
function.}

Consequently: {\it a two-dimensional Lorentzian metric is flat iff
the conformal factor factorizes in null coordinates}, that is to
say:
\[m(\texttt{u}, \texttt{v}) = U(\texttt{u}) V(\texttt{v}) \]
In this flat case, the change of coordinates from a generic null
system $\{\texttt{u},\texttt{v}\}$ to an inertial one
$\{\bar{\texttt{u}},\bar{\texttt{v}}\}$ is given by
\begin{equation} \label{nullinternal}
\begin{array}{l}
\bar{\texttt{u}}= \bar{\texttt{u}}(\texttt{u})  \\[1mm]
\bar{\texttt{v}}= \bar{\texttt{v}}(\texttt{v})
\end{array}
\end{equation}
where $\quad \bar{\texttt{u}}'(\texttt{u}) = U(\texttt{u}) \, ,
\quad \bar{\texttt{v}}'(\texttt{v}) = V(\texttt{v})$.

For a non (necessarily) flat metric, the change
(\ref{nullinternal}) gives the internal transformation between
null coordinates, the metric tensor changing as:
\begin{eqnarray*}
d s^2 = m(\texttt{u}, \texttt{v})\, d\texttt{u} \,
d\texttt{v} = \bar{m}(\bar{\texttt{u}}, \bar{\texttt{v}})\,
d\bar{\texttt{u}} \, d\bar{\texttt{v}}\\
m(\texttt{u}, \texttt{v}) = \bar{m}(\bar{\texttt{u}}(\texttt{u}),
\bar{\texttt{v}}(\texttt{v})) \, \bar{\texttt{u}}'(\texttt{u}) \,
\bar{\texttt{v}}' (\texttt{v})
\end{eqnarray*}

What does this internal transformation (\ref{nullinternal}) mean
from a geometric point of view? Any Lorentzian two-dimensional
metric defines in the corresponding space-time two (geodesic) null
congruences of curves or, what is two-dimensionally equivalent, two
families of null hypersurfaces. The space-time functions that have
one of these families as level hypersurfaces are the null coordinate
functions. Evidently, these functions are defined up to a change
given by (\ref{nullinternal}). So, the coordinate lines (or the
coordinate hypersurfaces) are invariant under this internal
transformation but they are parameterized in a different way in
every null coordinate system.

\section{Two-dimensional kinematics in null coordinates}
\label{ap-B}

Now we will consider some basic kinematic results expressed in a
given null coordinate system $\{\texttt{u},\texttt{v}\}$. In terms
of his proper time $\tau$, the trajectory of an observer $\gamma$
is:
\begin{equation} \label{observer}
\begin{array}{l}
\texttt{u}= u(\tau)  \\[1mm]
\texttt{v}= v(\tau)
\end{array}
\end{equation}
and its tangent vector:
\[T(\tau) = (T^{{\mbox{\scriptsize u}}}, T^{{\mbox{\scriptsize v}}}) =
(\dot{u}(\tau), \dot{v}(\tau))\]
where a dot means derivative with respect proper time. The unit
condition for $T$ connects the metric function $m(\texttt{u},
\texttt{v})$ with the observer trajectory (\ref{observer}):
\begin{equation} \label{B-unit}
m(u(\tau),v(\tau)) =\frac{1}{\dot{u}(\tau) \dot{v}(\tau)}
\end{equation}
This relation plays an important role in two-dimensional positioning
and states that: {\it when the unit tangent vector of an observer is
known in terms of his proper time, the metric on the trajectory of
this observer is also known}.

The proper time parameterized trajectory (\ref{observer})
tantamounts to a (geometric) trajectory $\texttt{v} = F(\texttt{u})$
and a proper time function $\tau = \tau(\texttt{u})$ restricted by
the unit condition. From one of the expressions (\ref{observer}) we
can obtain the proper time of the observer $\gamma$, say:
\[ \tau = \tau(\texttt{u}) = u^{-1}(\texttt{u})
\]
Then the trajectory is given by:
\[\texttt{v}=F(\texttt{u}) = v(\tau(\texttt{u})) \, ,
\]
and the unit condition (\ref{B-unit}) can be written:
\begin{equation} \label{trajectory}
[\tau'(\texttt{u})]^2 = m(\texttt{u},F(\texttt{u})) F'(\texttt{u})
\end{equation}

From expression (\ref{trajectory}) it follows that, when a proper
time relation $\tau = \tau(\texttt{u})$ is previously given, on any
fixed event a trajectory passes which admits this $\tau$ as proper
time. This is because to fix the event $(\texttt{u}_0,
\texttt{v}_0)$ is to fix an initial condition for $F({\rm u})$,
namely $\texttt{v}_0 = F(\texttt{u}_0)$. In other words, there
always exists a congruence of users having a prescribed proper time
function.

The {\it acceleration of the observer} (\ref{observer}) in null
coordinates $\{\texttt{u},\texttt{v}\}$ takes the expression:
\begin{equation*}
a(\tau) = ( a^{\mbox{\scriptsize u}}  , a^{\mbox{\scriptsize v}}
)= \left(\ddot{u} + (\ln m),_{\mbox{\scriptsize u}}\dot{u}^2, \,
\ddot{v} + (\ln m),_{\mbox{\scriptsize v}}\dot{v}^2 \right)
\end{equation*}
and the {\it acceleration scalar} $\alpha(\tau) \equiv \pm
\sqrt{-a^2(\tau)}$ is:
\begin{equation} \label{B-ac-scalar}
\alpha(\tau) = \frac{\ddot{u}}{\dot{u}} + (\ln
m),_{{\mbox{\scriptsize u}}}\dot{u} = -\frac{\ddot{v}}{\dot{v}} -
(\ln m),_{{\mbox{\scriptsize v}}}\dot{v}
\end{equation}

The {\it dynamic equation}, i.e. the equation for the world lines
with a known acceleration $\alpha$, and consequently the {\it
geodesic equation} (when $\alpha = 0$), can be written as two
coupled equations for the {\it proper time functions} $\, u(\tau)$
and $v(\tau)$:
\begin{equation}  \label{geodesic}
\begin{array}{c}
\displaystyle \frac{\ddot{u}}{\dot{u}} + (\ln m),_{u}\dot{u}  =
\alpha(\tau)  \\ \vspace{-4mm}  \\
m \dot{u} \dot{v}  =  1
\end{array}
\end{equation}
In (\ref{geodesic}) $m(\texttt{u},\texttt{v})$ is known, and $m$
stands for $m(u(\tau), v(\tau))$; therefore, it is a coupled system.

\section{Principal observers for two-dimensional null coordinates}
\label{ap-C}

In appendix B we have seen that every proper time function defines a
congruence of observers. To every null coordinate system
$\{\texttt{u}, \texttt{v}\}$ two special congruences of observers
exist, called {\it principal observers} of the system: those whose
proper time coincides with one of the two null coordinates. Thus, we
have the {\rm $\texttt{u}$}-{\em principal observers}:
\begin{equation} \label{u-principalobserver}
\begin{array}{l}
{\rm u}= \tau  \\
{\rm v}= \varphi_{{\rm u}} ({\rm u})
\end{array}
\end{equation}
and the {\rm $\texttt{v}$}-{\em principal observers}:
\begin{equation} \label{v-principalobserver}
\begin{array}{l}
{\rm u}= \varphi_{{\rm v}} ({\rm v})  \\
{\rm v}=\tau
\end{array}
\end{equation}

The tangent vectors to the principal observers are:
\[
T_{\rm u} = (1, \dot{\varphi}_{\rm u}) \, , \qquad   T_{\rm v} =
(\dot{\varphi}_{\rm v}, 1) \, ,
\]
and the unit condition takes the form:
\begin{equation} \label{C-unit}
m(\texttt{u}, \varphi_{\rm u}({\mbox u}))
 =
\frac{1}{\dot{\varphi}_{\rm u}({\rm u})} \, , \quad m(\varphi_{\rm
v}(\texttt{v}), \texttt{v}) = \frac{1}{\dot{\varphi}_{\rm
v}(\texttt{v})}
\end{equation}
For the principal observers the acceleration becomes:
\begin{equation}
\begin{array}{l}
a_{\rm u}({\mbox u}) = \left((\ln m),_{\rm u} \, , \,
\ddot{\varphi}_{\rm u} + (\ln
m),_{\rm v}\dot{\varphi}_{\rm u}^2 \right)  \\[2mm]
a_{\rm v}(\texttt{v}) = \left(\ddot{\varphi}_{\rm v} + (\ln
m),_{\rm u}\dot{\varphi}_{\rm v}^2 \, , \, (\ln m),_{\rm v}
\right)
\end{array}
\end{equation}
\vspace{1mm}\\
And the acceleration scalar is:
\begin{equation} \label{acc-scalar-po}
\begin{array}{lcl}
 \alpha_{\rm u}(\texttt{u}) & = & (\ln m),_{\rm
u}(\texttt{u}, \varphi_{\rm u}({\mbox{u}})) =  \\[1mm] & = & \displaystyle
-\frac{\ddot{\varphi}_{\rm u}(\texttt{u})}{\dot{\varphi}_{\rm
u}(\texttt{u})} - \dot{\varphi}_{\rm u}(\texttt{u})(\ln m),_{\rm
v}(\texttt{u}, \varphi_{\rm u}({\mbox{u}}))
      \\[4mm]
\alpha_{\rm v}(\texttt{v}) & = & - (\ln m),_{\rm
v}(\varphi_{\rm v}(\texttt{v}), \texttt{v}) =  \\[1mm]  & = & \displaystyle
\frac{\ddot{\varphi}_{\rm v}(\texttt{v})}{\dot{\varphi}_{\rm
v}(\texttt{v})} + \dot{\varphi}_{\rm v}(\texttt{v})(\ln m),_{\rm
u}(\varphi_{\rm v}(\texttt{v}), \texttt{v})
\end{array}
\end{equation}

\end{document}